\documentclass[11pt]{article} %
\usepackage{rldmsubmit,palatino}
\usepackage{graphicx}
\usepackage{amsmath}
\usepackage{xspace}
\usepackage{amssymb}
\usepackage{bm}
\usepackage{color}
\usepackage{url}
\usepackage{booktabs}
\usepackage{subfigure}
\usepackage{tikz}
\usetikzlibrary{positioning}

\usepackage[english]{babel}															%
\usepackage[protrusion=true,expansion=true]{microtype}
\usepackage{amsmath,amsfonts,amsthm} %
\usepackage{graphicx}
\usepackage{url}
\usepackage{hyperref}
\usepackage[boxed, linesnumbered]{algorithm2e}
\usepackage{listings}
\usepackage{verbatim}

\title{Skynet: A Top Deep RL Agent in the\\ Inaugural Pommerman Team Competition}

\author{Chao Gao\thanks{cgao3@ualberta.ca Department of Computing Science, University of Alberta. Work performed as intern at Borealis AI}, Pablo Hernandez-Leal, Bilal Kartal, and Matthew E. Taylor\\
Borealis AI\\
Edmonton, Alberta, Canada\\
\texttt{\{pablo.hernandez, bilal.kartal, matthew.taylor\}@borealisai.com} \\
}

\begin{document}

\maketitle

\begin{abstract}
The Pommerman Team Environment is a recently proposed benchmark which involves a multi-agent domain with challenges such as partial observability, decentralized execution (without communication), and very sparse and delayed rewards. The inaugural Pommerman Team Competition held at NeurIPS 2018 hosted 25 participants who submitted a team of 2 agents. Our submission \verb|nn_team_skynet955_skynet955| won 2nd place of the ``learning agents'' category.  Our team is composed of 2 neural networks trained with state of the art deep reinforcement learning algorithms and makes use of concepts like reward shaping, curriculum learning, and an automatic reasoning module for action pruning. Here, we describe these elements and additionally we present a collection of open-sourced agents that can be used for training and testing in the Pommerman environment. Code available at: \url{https://github.com/BorealisAI/pommerman-baseline}
\end{abstract}

\acknowledgements{}

\startmain %

\section{Introduction}

The \href{www.pommerman.com}{Pommerman} environment for benchmarking Multi-Agent Learning is based on the classic console game \textbf{Bomberman}. The team competition involves 4 bomber agents initially placed at the four corners of an $11$$\times$$11$ board. Each two diagonal agents form a team. At every step, each agent issues an action simultaneously from 6 discrete candidate moves: moving \verb|left|, \verb|right|, \verb|up|, \verb|down|,  placing a \verb|bomb|, or \verb|stop|.
The \verb|bomb| action is legal as long as the agent's \verb|ammo| is greater than 0, and any illegal action is superseded with \verb|stop| by the environment. Each cell on the board can either be a \verb|passage|, a \verb|rigid| wall, or \verb|wood|. Only \verb|passage| is passable for the agent. \verb|Wood| would be destroyed if it is covered by an exploding bomb's flame while a \verb|rigid| location is unaffected. Also, when \verb|wood| is destroyed a \verb|powerup| might appear (according to some probability) at the same location. There are 3 types of powerups: \verb|ExtraBomb|, \verb|EnableKick|, and \verb|ExtraBlast| for which the agent respectively, increases its ammo by 1, acquires the ability to kick bombs, and rises its bombs' blast strength by 1.

The game starts with a procedurally generated random map and each agent initially has an \verb|ammo| of 1 and \verb|blast strength| of 2. Whenever an agent places a bomb, it explodes after 10 time steps, producing flames that have a lifetime of 2 time steps and a radius of the bomb placing agent's \verb|blast strength|. The game ties when both teams have at least one agent alive after 800 steps. The team environment is also partially observable, meaning each agent can only see the local board with a radius of 4 cells, see Figure~\ref{fig:board}.

Pommerman is a challenging benchmark for multi-agent learning and model-free reinforcement learning, due to the following characteristics:

\textbf{Sparse and deceptive rewards}: the former refers to the fact that the only non-zero reward is obtained at the end of an episode. The latter refers to the fact that quite often a winning reward is due to the opponents' involuntary suicide, which makes reinforcing an agent's action based on such a reward \emph{deceptive}. Note that suicide happens frequently during learning since an agent has to place bombs to explode wood to move around on the board, while due to terrain constraints, in some cases, performing non-suicidal bomb placement requires complicated, long-term, and accurate planing.

\textbf{Delayed action effects}: the only way to make a change to the environment (e.g., bomb wood or kill an agent) is by means of bomb placement, but the effect of such an action is only observed when the bomb's timer decreases to 0; more complications are added when a placed bomb is kicked to another position by some other agent.

 \textbf{Imperfect information}: an agent can only see its nearby areas. This makes the effect of certain actions, such as bomb kicking, unpredictable. Indeed, even detecting which agent placed the exploding bomb is intractable in general because of the hidden information of the board.

\textbf{Uninformative multiagent credit assignment}: In the team environment, the same episodic reward is given to two members of the team. It may not be clear how to assign credit to individual agents. For example, consider an episode where an agent eliminates an opponent but then commits suicide, and its teammate eliminates the remaining opponent. Under this scenario, both team members get a positive reward from the environment, but this could reinforce the suicidal behaviour of the first agent. Similarly, one agent could eliminate both opponents whereas its teammate just camps; both agents would get positive rewards, reinforcing a \emph{lazy} agent~\cite{Panait:2005wj}.

\section{Skynet955}
\verb|nn_team_skynet955_skynet955| is a team composed of two identical neural networks, where
\verb|skynet955| is trained after equipping the neural net agent with an ``ActionFilter'' module for 955 iterations. The philosophy is to instill prior knowledge to the agent by telling the agent \emph{what not to do} and let the agent discover \emph{what to do} by trial-and-error. The benefit is twofold: 1) the learning problem is simplified since suicidal actions are removed and bomb placing becomes safe; and 2) superficial skills such as not moving into flames and evading bombs in simple cases are handled.
Below we describe the main components of our team: the ActionFilter and the reinforcement learning aspect.

\begin{figure}
\centering
\subfigure[]{
\includegraphics[scale=0.10]{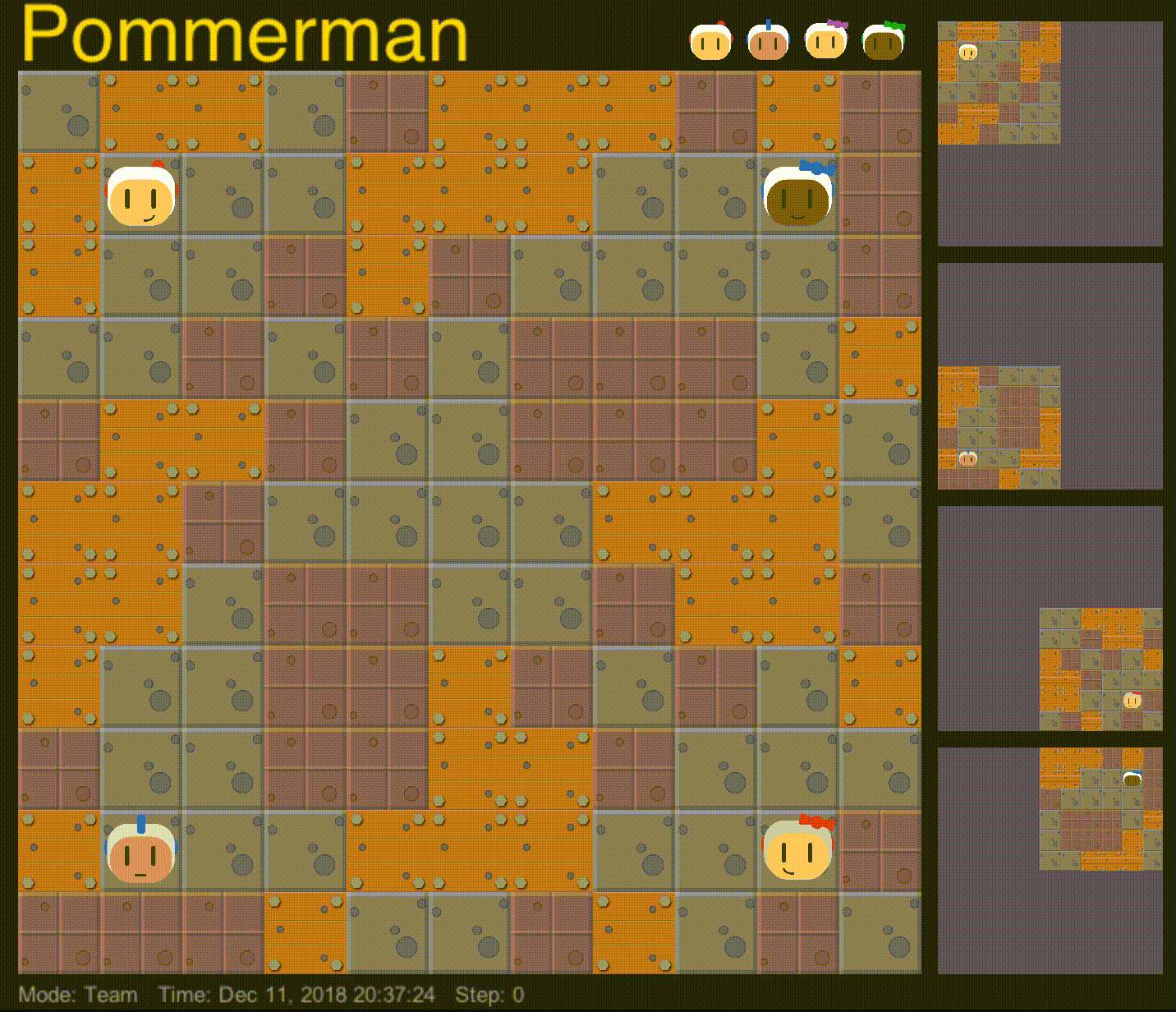}
\label{fig:board}
}
\subfigure[]{
\includegraphics[width=0.6\textwidth]{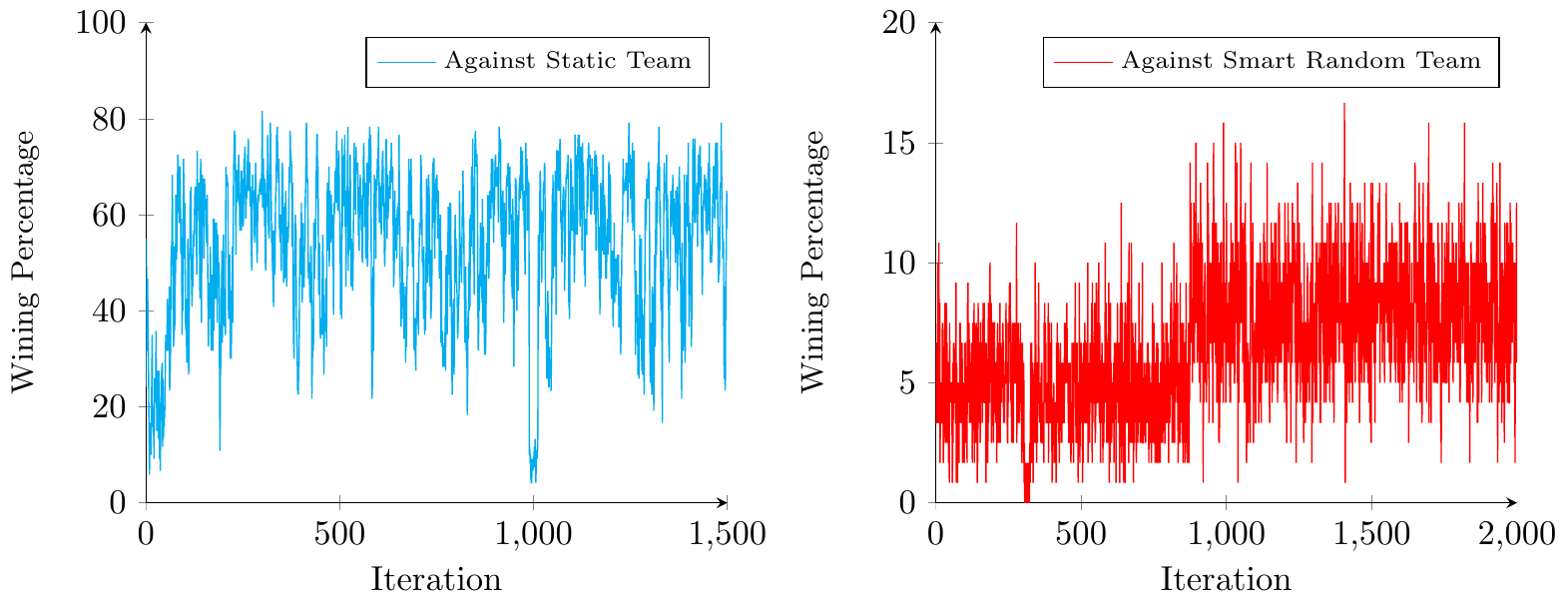}
\label{fig:curves}
}
\caption{(a) An example team environment Pommerman game. On the right is shown the partial observation (a limited area around) for each agent.  (b) Winning percentage of the learning team against a team of \texttt{Static} opponents and a team of \texttt{Smart} random opponents.}
\label{fig:1}
\end{figure}

\paragraph{ActionFilter}
\label{sec:actionFilter}

We designed the filter to speed up learning so that agents can focus on higher level strategies. The filter thus serves as a safety check to provide more efficient exploration. The ActionFilter is implemented by rules  shown in Table \ref{tab:rules}. Note that for the ``avoiding suicide'' rules, \verb|skynet955| implemented a simple version of them (e.g., a moving bomb was simply treated as static); a full implementation would arguably make the agent stronger. It is worth mentioning that the above ActionFilter is extremely fast. In our experiments, together with neural net evaluation, each action takes $\approx$ 3 ms on a GTX 1060 GPU on average, while the time limit in the competition is 100 ms per move. Also, we note that another natural approach for ``bomb placement'' pruning is conducting a lookahead search using ``avoding suicide'' rules; this is perhaps better than the crude rules described above. We think this ActionFilter will be useful for learning agents within the Pommerman domain, since it can constrain the action space during learning without significantly affecting the overall team strategy. Ideally, one can even start model-free learning with the filter, and once an agent acquires some basic skills, it can be unplugged from the neural network so that strategy bias by filter is completely removed. For this reason, we have open-sourced an implementation of the ActionFilter.\footnote{ \url{https://github.com/BorealisAI/pommerman-baseline}}

\begin{table}[]
\small
    \centering
     \caption{ActionFilter rules}
    \begin{tabular}{@{}p{2.5cm}p{15.5cm}@{}}
    \toprule
     \emph{Avoiding Suicide} & Not going to positions that are flames on the next step. \\
       & Not going to \emph{doomed} positions, i.e., positions where if the agent were to go there the agent would have no way to escape. For any bomb, doomed positions can be computed by referring to its \texttt{blast strength},  \texttt{blast range}, and \texttt{life}, together with the local terrain.\\
    \emph{Bomb Placement}&  Not place bombs when teammate is close, i.e., when their Manhattan distance is less than their combined blast strength.  \\
		&Not place bombs when the agent's position is covered by the blast of any previously placed bomb. \\
	\bottomrule
    \end{tabular}
    \label{tab:rules}
\end{table}

\begin{figure}
\centering
\includegraphics[scale=1.0]{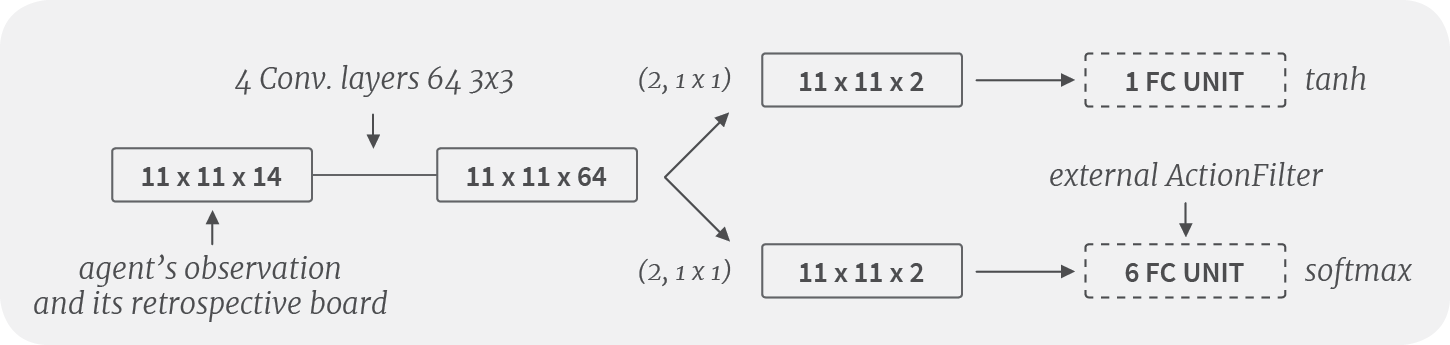}
\label{fig:arch}
\caption{Architecture used for the \texttt{skynet955} agents}
\end{figure}

\paragraph{Reinforcement Learning}
As depicted in Figure~\ref{fig:arch}, the architecture contains 4 convolution layers, followed by policy and value heads. The input contains 14 features planes, each of shape $11$$\times$$11$, similar to~\cite{resnick2018pommerman}. It then convolves using $4$ layers of convolution, each has 64 $3$$\times$$3$ kernels; the result thus has shape $11$$\times$$11$$\times$$64$. Then, each head convolves using $2$ $1$$\times$$1$ kernels. Finally, the output is squashed into action probability distribution and value estimation, respectively. Such a two-head architecture is a natural choice for Action-Critic algorithms, as it is generally believed that forcing policy and value learning to use shared weights could reduce over-fitting~\cite{schulman2017proximal}.

Instead of using an LSTM to track the history observations, we use a ``retrospective board'' to remember the most recent value at each cell of the board, i.e., for cells outside of an agent's view, in the ``retrospective board'' its value is filled with what was present when the agent saw that cell the last time. The input feature has 14 planes in total, where the first 10 are extracted from the agent's current observation, the rest 4 are from ``retrospective board.'' We initially performed experiments with an LSTM to track all previous observations; however, due to computational overhead on top of the already prolonged training in Pommerman domain, we decided to replace it with the ``retrospective board'' idea, which yielded similar performance but was significantly faster.

The neural net is trained by \emph{PPO}~\cite{schulman2017proximal}, minimizing the following objective:
\begin{equation}
    o(\theta;\mathcal{D})  = \sum_{(s_t, a_t, R_t) \in \mathcal{D}} \Bigg[ -\mathit{clip}(\frac{\pi_\theta(a_t|s_t)}{\pi_\theta^{old}(a_t|s_t)}, 1-\epsilon, 1+\epsilon) A(s_t, a_t) +
     \frac{\alpha}{2} \cdot \max\Big[ (v_\theta(s_t) -R_t)^2, (v_\theta^{old}(s_t) + \mathit{clip}(v_\theta(s_t) - v_\theta^{old}(s_t), -\epsilon, \epsilon)-R_t)^2 \Big] \Bigg],
\end{equation}
where $\theta$ is the neural net, $\mathcal{D}$ is sampled by $\pi_\theta^{old}$, and $\epsilon$ is a tuning parameter. Refer to OpenAI baseline for details~\cite{schulman2017proximal}.

\paragraph{Curriculum Learning}
Training is conducted by letting two identical neural net players compete against a set of curriculum~\cite{bengio2009curriculum} opponents: (i) \verb|Static| opponent teams, where opponents do not move or place bombs. Competing against a team of \verb|Static| opponents teaches our agents to get closer to opponents, place a bomb, and move away to a safe zone.  The trained neural net is then used against the second opponent in the curriculum. (ii) \verb|SmartRandomNoBomb|: players that do not place bombs. \emph{Smart} random means it has the ActionFilter as described earlier and the action taken is random (except that bomb placing is disallowed). The reason we let \verb|SmartRandomNoBomb| not place bombs is that the neural net can focus on learning true ``killing'' skills, not a skill that solely relies on the opponent's strategy flaw (e.g., the provided baseline \verb|SimpleAgent| has a significant flaw where the competitor can diagonally block and make \verb|SimpleAgent| be killed by its own bomb). This avoids the ``false positive'' reward signal caused by opponent's involuntary suicide. Competing against a team of \verb|SmartRandomNoBomb| helps our agents to learn better battling skills such as using the topological map to corner the opponents and pursuing opponents.

\begin{table}[]
    \centering
    \small
    \caption{Reward Shaping for \texttt{skynet955} agents}
    \begin{tabular}{@{}p{9cm}p{9cm}@{}}
    \toprule
    Going to a cell not in a 121-length FIFO queue gets $0.001$. &
    At the end of a game, dead agent in the winning team gets $0.5$.  \\
    Picking up \verb|kick| gets $0.02$. & For draw games, all agents receive $0.0$. \\
    Picking up \verb|ammo| gets $0.01$. & On one enemy's death gets $0.5$. \\
    Picking up \verb|blast strength| gets $0.01$. & On a teammate's death gets $-0.5$.  \\
  	\bottomrule
    \end{tabular}
    \label{tab:shaping}
\end{table}

\textbf{Reward shaping:} To cope with the sparse reward problem, a dense reward function is added during the learning, see Table~\ref{tab:shaping}. It should be noted the above hand designed reward function is still noisy in the sense that an agent's contribution was not clearly separated.

\textbf{Results:} Figure~\ref{fig:curves} shows the learning curves against \verb|Static| and \verb|SmartRandomNoBomb| teams. In our training, each iteration contains 120 games, produced in parallel by 12 actor workers. The curves show that, against \verb|Static| agents, the neural net achieved wining percentage around 70\%, while against \verb|SmartRandomNoBomb|, it never reached 20\%. %
We note that because the opponents do not place bombs, the rest of the games are almost all draws. The learning seems to be slow, in part because playing against \verb|SmartRandomNoBomb|, a large number of games were ended with draws, which gives reward signal $0$ in our training.

In the competition, our team was composed of two identical neural net models, at each step, for each of our agent, each action typically costs one to several milliseconds, while the time limit is 100ms per move. The submitted agent \verb|skynet955| was the neural net model at iteration $955$ obtained in training against \verb|SmartRandomNoBomb| team, as by the time of submission only around 1000 iterations were finished.
Throughout our training and testing, only the \verb|V0| environment (which has no wall collapsing) was used. By contrast, in the competition, the \verb|V1| (which has wall collapsing, meaning at certain time step, the boarder passages will suddenly change to rigid walls, and any agent that happens to be in any of the corresponding cells dies) was used.

\paragraph{An open-sourced collection of agents for Pommerman}

\begin{table}[]
    \centering
     \caption{Description of different agents. Our open-sourced agents: \texttt{StaticAgent, SmartRandom, SmartRandomNoBomb} and \texttt{CautiousAgent}, will be helpful to baseline against and to train against. The bottom 3 agents are relatively stronger than the provided baseline, less prone to exploitation due to higher level of stochasticity, and fast decision makers to be used during training.}
    \begin{tabular}{@{}p{6.0cm}p{12.2cm}@{}}
    \toprule
    \verb|SimpleAgent| (Provided by Pommerman) & A heuristic agent that uses Dijkstra and rules for navigation, tune-up collection, and simple attacks.\\
     \hline
    \verb|StaticAgent| & A boring agent that always executes the stop action. Helpful for learning agents as rewards are noise-free. \\
    \hline
    \verb|SmartRandom| (Random agent + ActionFilter) & Takes random actions from the filtered action space. Stochastic but careful actions render it a competitive opponent for RL agents. \\
    \hline
    \verb|SmartRandomNoBomb| (Random agent + ActionFilter + No Bombing) & Similar to \verb|SmartRandom| but force the agent not place bomb at all. \\
    \hline
    \verb|CautiousAgent| (Modified  SimpleAgent) & The modification enables the agent to place a bomb if and only if it guarantees a kill.\\
  	\hline
    \verb|Skynet955| (Neural Network agent) & Agent that is trained with PPO using reward shaping, ActionFilter, and opponent curriculum learning.\\
  	\bottomrule
    \end{tabular}
    \label{tab:agents}
\end{table}

 Against \verb|SimpleAgent|, it is not difficult to train a non-placing-bomb neural net agent that wins by diagonally blocking and forcing \verb|SimpleAgent| to get stuck on its self-placed bomb.\footnote{\url{https://youtu.be/3yUhI46Xx8o}} This strategy flaw of \verb|SimpleAgent| stems from its hand-crafted heuristic strategy for enemy engagement and bomb placement. Learning agents exploit this flaw and the learned policy does not generalize against other opponents types. Therefore, we propose and open source different agents to be used for training, see Table \ref{tab:agents}.

\verb|StaticAgent| is an extremely simple agent that always executes the stop action, the advantage of using this agents is that rewards are noise-free. \verb|SmartRandomNoBomb| agent is a very challenging opponent. It moves randomly among the filtered actions, therefore learning against it provides better generalization. It does not place bombs, and thus never commits suicide. In contrast to \verb|SimpleAgent| and \verb|CautiousAgent|, it does not use Dijkstra, and it takes less time to act.

\verb|CautiousAgent| is based on a modification of  \verb|SimpleAgent|, the main idea is to only let this agent place a bomb when it is certain to kill an opponent. However, with this adaptation, the ``weak'' opponent \verb|SimpleAgent| is instantly turned into a ``strong'' player. Killing this agent requires quite advanced skills. When learning against this opponent, a large number of games ended with draws, or the challenger is killed. For example, in our tests, our \verb|Skynet| agent can easily achieve 70\% wining percentage against \verb|SimpleAgent| team, but only around 10\% against \verb|CautiousTeam|, if not excluding draws.

\section{Conclusions and Future work}

In spite of the good performance of our \verb|skynet| agents, there are still many potential avenues for future research. For example, recent innovations, such as curiosity~\cite{burda2018exploration,pathak2017curiosity}, centralized learning with decentralized execution~\cite{Foerster:2017ti}, or difference rewards~\cite{Foerster:2017uq} have shown to be able to produce good results in related domains, it remains to experimentally verify how effective they would be in the challenging domain of Pommerman.

One challenge of multi-agent learning in partial observable environments is the credit assignment for each individual agent's behavior~\cite{devlin2014potential,hernandez2018multiagent}. Whenever there is a success or failure, it is important to correctly identify who is to reward or blame. Unfortunately, the original environment provided by Pommerman does not provide any such information. However, if centralized training~\cite{Foerster:2017ti} can be used by revising the environment, information might be helpful in devising more accurate reward function, for example: identification of bombs' owners and  bombs' kickers; in general for any relevant event occurred (wood destruction, enemy's death, etc.) identifying which agent is to reward or blame could be based on who is responsible for the corresponding exploding bomb. Lastly, an ensemble consists of multiple neural net models may also improve the playing performance in the competition setting.

\small
 \setlength{\parskip}{0pt}
  \setlength{\itemsep}{0pt}
\bibliographystyle{abbrv}
\bibliography{bib}

\end{document}